# Unveiling the *ac* Dynamics of Ferroelectric Domains by Investigating the Frequency Dependence of Hysteresis Loops

S. M. Yang[1], J. Y. Jo[1], T. H. Kim[1], J.-G. Yoon[2], T. K. Song[3], H. N. Lee[4], Z. Marton[4,5], S. Park[6], Y. Jo[6], and T. W. Noh[1,*]

[1]ReCOE, Department of Physics and Astronomy, Seoul National University, Seoul 151-747, Korea

[2]Department of Physics, University of Suwon, Hwaseong, Gyunggi-do 445-743, Korea

[3]School of Nano and Advanced Materials Engineering, Changwon National University, Changwon, Gyeongnam 641-773, Korea

[4]Materials Science and Technology Division, Oak Ridge National Laboratory, Oak Ridge, Tennessee 37831, USA

[5]Department of Materials Science and Engineering, University of Pennsylvania, Philadelphia, Pennsylvania 19104, USA

[6]Nanomaterials Research Team, Korea Basic Science Institute, Daejeon 305-333, Korea

We investigated nonequilibrium domain wall dynamics under an *ac* field by measuring the hysteresis loops of epitaxial ferroelectric capacitors at various frequencies and temperatures. Polarization switching is induced mostly by thermally activated creep motion at lower frequencies, and by viscous flow motion at higher frequencies. The dynamic crossover between the creep and flow regimes unveils two frequency-dependent scaling regions of hysteresis loops. Based on these findings, we constructed a dynamic phase diagram for hysteretic ferroelectric domain dynamics in the presence of *ac* fields.

PACS numbers: 77.80.Dj, 77.80.Fm, 05.45.-a

*Author to whom correspondence should be addressed: twnoh@snu.ac.kr

Hysteresis is an intriguing memory effect associated with dynamical responses driven by conjugating external forces. It occurs in numerous physical phenomena, including ferroelectricity, ferromagnetism, and superconductivity [1]. It also appears in chemistry [2], biology [3], and even in economics [4]. Although hysteresis phenomena were already known in the mid 1800s [5], several key issues are still poorly understood. One such fundamental issue is the frequency ($f$)-dependence of hysteresis loops: namely, how do the dynamic responses depend on the sweep rate of external *ac* force? Nowadays, understanding of the $f$-dependence of hysteresis loops has become technologically important because many nanoscale ferroic devices, such as non-volatile memories and data storage, require operations at high $f$.

In this respect, uniaxial ferroelectric (FE) materials with two metastable states separated by an energy barrier are of special interest. Each state is characterized by a scalar order parameter, i.e., the polarization $P$, which can be reversed by applying an electric field $E$. Ideally, if all spatial regions experience the $P$ reversal simultaneously at the same $E$ (called intrinsic switching [6]), the hysteresis loop would be $f$-independent. In reality, the free energy of a FE system prefers a configuration consisting of regions, called domains, each having the same $P$ direction. Under an external $E$, the volume of the favored FE domain state will increase by the domain nucleation and subsequent domain wall (DW) motion [7]. Due to the interaction between DWs and defects, the DW should experience time-dependent viscous motion, resulting in the $f$-dependence of the hysteresis loops under an *ac* field. However, there has been little effort to understand the $f$-dependence of FE hysteresis loops in terms of *ac* field-induced nonequilibrium DW dynamics.

It is well-known that the FE DW motion can be treated as the pinning-dominated driven dynamics of an elastic object in a random environment [8-12]. In a real FE medium, elastic forces tend to keep DWs flat, while pinning forces locally promote wandering. This competition leads to a complicated energy landscape with many local minima. That is, the FE DW velocity $v$ has various dynamic regimes, including pinned, thermally activated creep, and viscous flow, depending on the magnitude of external $E$ [8, 9]. Note that this concept has been used as a model to explain numerous physical systems, including charge density waves [13], contact lines in wetting [14], and DWs of ferromagnetic systems [8, 15, 16]. Therefore, investigation of the $f$-dependent $P$-$E$ hysteresis loops of FE systems could provide us not only with an understanding of the FE DW dynamics, but also with new insights into other related *ac* field-induced physical phenomena.

In this Letter, we report our investigation into how the nonequilibrium DW dynamics under an *ac* field affect the $f$-dependence of the hysteresis loops of a uniaxial FE system. We measured the $P$-$E$ hysteresis loops of a high-quality FE PbZr$_{0.2}$Ti$_{0.8}$O$_3$ (PZT) thin film at various $f$ and temperatures ($T$). We found two scaling regions for the $f$-dependence of coercive field $E_C$ for all measured $T$. We show that this $f$-dependence of $E_C$ is related to the dynamic crossover between the creep and viscous flow regimes of DW motions. Based on these experimental findings, we constructed a dynamic phase diagram for hysteretic domain dynamics under an *ac* field. This work provides a bridge between the $f$-dependence of hysteresis loops and the nonequilibrium *ac* dynamics of DWs in random media.

We used high-quality FE capacitors, consisting of a 100-nm-thick epitaxial PZT film deposited on a SrRuO$_3$/SrTiO$_3$ (001) substrate with Pt top electrodes. A more detailed description of the growth procedures has been published elsewhere [17]. X-ray diffraction studies confirmed that it was composed of purely *c*-axis oriented domains, i.e., the film had uniaxial symmetry and 180° DW configurations. We measured $P$-$E$ hysteresis loops under various triangular wave $f$ (5 - 2000 Hz) and $T$ (10 - 300 K).

The detailed shape of the $P$-$E$ hysteresis loop depended on $f$. Figure 1(a) shows the $f$-dependent hysteresis loops for the PZT film at 300 K. At the lowest $f$, the hysteresis loop looked almost like a square. As $f$ increased, $E_C$ increased and the slope at $E_C$ decreased without any other significant changes. This indicates that the DW dynamics around $E_C$ plays a major role in the $f$-dependent shape change of hysteresis loops.

We found that $T$ could also significantly affect the shape change of the $P$-$E$ hysteresis loops. Figure 1(b) shows $T$-dependent hysteresis loops for the PZT film at $f$ = 1000 Hz. To avoid complications that might come from a path-dependence in the hysteresis measurements, we always used the amplitude $E_{max}$ of a triangular wave at least 2.5 times larger than the corresponding value of $E_C$ for the given $T$. Contrary to the $f$-dependence, the slope at $E_C$ did not change significantly. However, as $T$ increased, there were distinctive decreases in $E_C$, indicating the involvement of thermally activated processes in the $P$ reversal.

To gain further insights into the *ac* DW dynamics, we looked into the $f$-dependence of $E_C$ at various $T$. As shown in Fig. 1(c), $E_C$ increased with $f$. The phenomenological power law relationship, $E_C \propto f^{\beta}$ [18], described our experimental $E_C$ values quite well for all $T$, as represented by the solid lines. Interestingly, we found two scaling regions with different $\beta$ values and a crossover frequency $f_{cr}$ = 200 - 500 Hz. We argue that these intriguing results are closely related to the nonequilibrium DW dynamics under an *ac* field.

When we apply an *ac* field with $E_{max} \gg E_C$, as shown in Fig. 2(a), the $E$ value will vary in time ($t$) during one period. Figure 2(b) shows a schematic diagram of how the DW motions will change with external $E(t)$ in energy landscapes $\phi(x)$ when initially DWs exist. As shown by the solid line, there is a wide distribution in energy barrier heights $U_B$ between local minima. At finite $T$, thermally activated hopping, known as creep motion [10], induced DW movement from one local minimum to the next. This process requires a certain hopping time $\tau$, which depends on the effective $U_B(E)$. When $E$ is very small, the effective $U_B(E)$ is quite large and so $\tau$ becomes very long. Then it is impossible to overcome $U_B$ during one *ac* field cycle (i.e., $\tau > 1/f$), there will be no macroscopic DW motion with $v = 0$, i.e., a *pinned regime* (purple arrow).

Conversely, when DWs do not initially exist (i.e., reversed domains are not generated), we should first consider a domain nucleation process. Domain nucleation with opposite $P$ will occur at a certain $E$ value, $E_N$. In the epitaxially grown film with uniaxial symmetry, generally, $E_N$ is larger than the $E$ value required for depinning process, $E_{depin}$ (i.e., $f\tau \sim 1$) [19]. In this case, there is a *no nuclei regime* instead of a pinned regime. The schematic diagram for this case is shown in Fig. 2(c).

For both cases, as shown in Figs. 2(b) and 2(c), there should be three regimes for FE DW motions in one *ac* cycle. (1) At $E < E_1$, the motion will be either in the *no nuclei* or *pinned* regime. (2) At $E_1 < E < E_2$, the effective $U_B$ is reduced and $\tau$ becomes shorter, so the DW starts to move by *thermally activated creep motion* (orange arrows). (3) At $E > E_2$, the effective $U_B$ almost vanishes, so the DW undergoes a *viscous flow motion* (magenta arrow). Therefore, $E_1$ and $E_2$ correspond to the initiation of creep motion and the dynamic crossover between the creep and flow regimes, respectively.

Depending on $f$, the relative time duration that the FE system stays in each regime will vary. Figure 2(d) shows the polarization reversal ($\Delta P$) during the first $1/4f$. For low $f$, as represented by the solid (blue) line, the system will stay in the creep regime long enough to induce most of the $\Delta P$. Since $E_C$ is the $E$ value at which 50 % of $\Delta P$ occur, $E_C < E_2$ for low $f$. Then all of the $\Delta P$ should occur in a rather narrow region of $E$, which results in the nearly square-like hysteresis observed in Fig. 1(a). On the other hand, for high $f$, $E$ increases quickly in time, so the system will only remain in the creep regime for a rather short time and then transit to the flow regime, as shown by the dotted (red) line. In this case, the DW motions in the flow regime will induce most of the $\Delta P$, so $E_C > E_2$. Since a rather large $E$ region is required to complete $\Delta P$, the hysteresis loop should be slanted, as shown in Fig. 1(a).

We therefore argue that the difference in contribution of DW motion depending on $f$ results in the two $E_C$ scaling regions observed in Fig. 1(c). As shown by the dashed (green) line, let us choose $f_{cr}$ as the frequency when $E_C$ becomes the same as $E_2$, i.e., $E_2 = E_C(f_{cr})$. Then if $f < f_{cr}$, the DW creep motion contributes most of the $\Delta P$. On the other hand, if $f > f_{cr}$, the flow motion will contribute most of the $\Delta P$.

Nattermann, Pokrovsky, and Vinokur developed a unified theory for thermally activated and flow DW dynamics in impure ferromagnetic systems [15]. Since DW dynamics of both FE and ferromagnetic systems can be treated as nonequilibrium dynamics of elastic objects in random media, we can naturally extend it to our FE systems. Let us consider a real DW segment, where the $U_B$ varies from place to place. The creep motion should be governed by the maximum energy barrier $U_{B,max}$. The time scale $\tau$ required to overcome $U_{B,max}$ due to thermal activation is of order [15]

$$\tau \approx \tau_0 \exp(U_{B,\max} / k_B T); \quad (1)$$
$$U_{B,\max} \approx U(E_{C0} / E)^\mu (1 - E / E_{C0})^\eta, \quad (2)$$

where $\tau_0$, $U$, and $E_{C0}$ are a microscopic hopping time, a characteristic pinning energy, and a depinning threshold field at $T = 0$, respectively. $\mu$ is a dynamical exponent for creep motion, and $\eta$ is an exponent related to the $E$-dependence of the energy barrier [20]. By evaluating the average velocity of the creep regime and comparing with $v \approx E$ [15], we obtain the $E_2$ value where the dynamic crossover from creep to flow regimes occurs:

$$E_2 / E_{C0} = \left[ (U / k_B T)(1 - E_2 / E_{C0})^\eta \right]^{1/\mu}. \quad (3)$$

Note that $E_2$ is a monotonously decreasing function of $T$ with a maximum $E_{C0}$ at $T = 0$.

To check the validity of our argument that $E_2 = E_C(f_{cr})$, we compared the experimental $E_C(f_{cr})$ values with Eq. (3). The symbols in Fig. 3(a) show the experimental values of $E_C(f_{cr})$ as a function of $T$. As $T$ increased, $E_C(f_{cr})$ decreased. It is known that $\eta = 2$ at $T = 0$ [20]. For similarly prepared PZT capacitors, our earlier *dc* DW dynamics study showed that $E_{C0} \sim 1.0$ MV/cm, $U/k_B \sim 300$ K, and $\mu \sim 0.9$ [9]. Thus, without any free parameter, Eq. (3) gives the solid line in Fig. 3(a), providing good agreement with the experimental $E_C(f_{cr})$ values below 50 K. At finite $T$, $\eta$ is known to be non-universal [15]. Thus, for the high $T$ region (100 - 300 K), we did another fit using $\eta$ as a free fitting parameter. The higher $T$ data fit quite well with $\eta = 4.0 \pm 0.5$ (dashed line). These agreements validate our argument that $E_2 = E_C(f_{cr})$, and the *f*-dependence of $E_C$ originates from the dynamic crossover between the creep and flow regimes.

Another crossover, $E_1$, determines the initiation of the thermally activated creep motion. We should determine whether there should be a *no nuclei* or a *pinned* regime below $E_1$. In other words, which schematic diagram is valid between Fig. 2(b) or 2(c)? As shown in Fig. 2(d), $E_1$ could be determined by the value where $\Delta P$ increases rapidly. Thus, we measured $E_1$ as the $E$ value where the switching current $I$ started to increase significantly, which is marked by the arrow in the inset of Fig. 1(a). The symbols in Fig. 3(b) display the experimental values of $E_1$ at various $T$.

If $E_1$ is governed by a depinning process of the DWs, as described in Fig. 2(b), we can estimate $E_1$ from $E_{depin}$, the required minimum field for depinning process. By using Eqs. (1) - (3), the $E_{depin}$ can be expressed as [15]

$$E_{depin}/E_{C0} = \left[(U/k_B T \Lambda)\left(1 - E_{depin}/E_{C0}\right)^{\eta}\right]^{1/\mu}. \qquad (4)$$

where $\Lambda = \ln(1/f\tau_0)$. Using the value of $1/\tau_0 \sim 10^{13}$ Hz, i.e., the attempt frequency for a typical phonon, we estimated the predicted $E_{depin}$ values. As shown by the solid (purple) region in Fig. 3(b), our measured $E_1$ values cannot be explained by the $E_{depin}$ values.

Therefore $E_1$ should result from the domain nucleation process described in Fig. 2(c). To create the reversed domain nuclei, the effective barrier height for nucleation, $U_n$ must be overcome also by thermal activation [21]. By substituting $U_n$ for $U$ in Eq. (4), we obtained a good fit for the $E_1$ values, as shown by the dashed (magenta) region in Fig. 3(b). We found $U_n/k_B \sim 1800$ K, which is about six times larger than $U$. Since our PZT film was epitaxially grown and composed of only *c*-domains, $U_n$ could be higher than $U$. The obtained $U_n$ value is consistent with the recent piezoresponse force microscope study on the energy distribution of nucleation centers in PZT films [22].

From these experimental findings, we constructed the dynamic phase diagram for domain dynamics under an *ac* field shown in Fig. 3(c). The (red) circles and (blue) squares represent experimental $E_1$ and $E_2$ values, respectively. Note that $E_1$ should be governed by the nucleation process and $E_2$ should correspond to the dynamic crossover from the creep to viscous flow regimes. The horizontal (green) line indicates the first $1/4f$ of $E(t)$ during the *P-E* hysteresis loop measurements. Depending on how fast it moves, the $\Delta P$ of the PZT capacitors is achieved mainly by either the creep or flow motion of the DWs, as shown in Fig. 2(d). This difference in dominant DW dynamics determines the *f*-dependence of hysteresis loops. The dynamic phase diagram provides us valuable information on the *f*- and *T*-dependence of hysteresis loops, such as the starting field for *P* reversal and the range of $E_C$ at the given *f* and *T*. We expect that this work can be generalized to other FE material systems, such as films with more than one type of domain, or polycrystalline FE films.

This research was supported by Basic Science Research Program through the National Research Foundation of Korea (NRF) funded by the Ministry of Education, Science and Technology (No. 2009-0080567). S. M. Y. acknowledges the financial support, in part, from a Seoul Science Fellowship. The work at Oak Ridge National Laboratory (HNL) was sponsored by the Division of Materials Sciences and Engineering, Office of Basic Energy Sciences, U.S. Department of Energy.

References

[1] A. Visintin, *Differential Models of Hysteresis* (Springer, Berlin, 1994).

[2] B. Xiao *et al.*, Nat. Chem. **1**, 289 (2009).

[3] J. R. Pomerening, E. D. Sontag, and J. E. Ferrell, Nat. Cell Biol. **5**, 346 (2003).

[4] W. Franz, Empirical Economics **15**, 109 (1990).

[5] J. C. Maxwell, *A treatise on electricity and magnetism* (Clarendon Press, Oxford, 1873).

[6] G. Vizdrik *et al.*, Phys. Rev. B **68**, 094113 (2003).

[7] M. Dawber, K. M. Rabe, and J. F. Scott, Rev. Mod. Phys. **77**, 1083 (2005).

[8] W. Kleemann, Annu. Rev. Mater. Res. **37**, 415 (2007).

[9] J. Y. Jo *et al.*, Phys. Rev. Lett. **102**, 045701 (2009).

[10] T. Tybell *et al.*, Phys. Rev. Lett. **89** (2002).

[11] P. Paruch, T. Giamarchi, and J. M. Triscone, Phys. Rev. Lett. **94** (2005).

[12] S. M. Yang *et al.*, Appl. Phys. Lett. **92**, 252901 (2008).

[13] G. Gruner, Rev. Mod. Phys. **60**, 1129 (1988).

[14] P. G. de Gennes, Rev. Mod. Phys. **57**, 827 (1985).

[15] T. Nattermann, V. Pokrovsky, and V. M. Vinokur, Phys. Rev. Lett. **87**, 197005 (2001).

[16] S. Brazovskii and T. Nattermann, Adv. Phys. **53**, 177 (2004).

[17] H. N. Lee *et al.*, Phys. Rev. Lett. **98**, 217602 (2007).

[18] Y. Ishibashi and H. Orihara, Integr. Ferroelectr. **9**, 57 (1995).

[19] S. M. Yang *et al.*, J. Korean Phys. Soc. **55**, 820 (2009).

[20] A. A. Middleton, Phys. Rev. B **45**, 9465 (1992).

[21] J. Y. Jo *et al.*, Phys. Rev. Lett. **97**, 247602 (2006).

[22] S. Jesse *et al.*, Nat. Mater. **7**, 209 (2008).

Figure captions

FIG. 1 (color online). (a) $f$-dependent $P$-$E$ hysteresis loops of the PZT film at 300 K. The inset shows the switching current $I$-$E$ curve at $f=$ 1000 Hz. (b) $T$-dependent $P$-$E$ hysteresis loops at $f=$ 1000 Hz. (c) $E_C$ values as a function of $f$ at various $T$. The $f_{cr}$ indicates the crossover frequency where two scaling regions are separated. The solid lines show the linear fitting results for each $f$ region.

FIG. 2 (color online). (a) Schematic diagram of triangular waves of low $f$ (solid blue line) and high $f$ (dotted red line) used for measuring hysteresis loops. Schematic diagram of energy landscapes $\phi(x)$ (solid line) and various DW motions in random media during the first $1/4f$, (b) when DWs exist initially and (c) when they do not exist initially. (d) Schematic diagram of $\Delta P/2P_r$ as a function of $E/E_{max}$, where $P_r$ is the remnant polarization. The dark gray, gray, and white regions in (a) and (d) indicate the pinned or no nuclei, creep, and flow regimes, respectively. The dashed (green) line in (d) shows the $E$-dependent $\Delta P/2P_r$ value at $f=f_{cr}$.

FIG. 3 (color online). (a) Experimental values of $E_C(f_{cr})$ as a function of $T$. The solid and dashed lines indicate the fitting results using Eq. (3) with $\eta=$ 2.0 and about 4.0, respectively. (b) Experimental values of $E_1$ as a function of $T$. The solid (purple) and dashed (magenta) regions indicate the expected $E_1$ values by assuming the depinning process and domain nucleation process, respectively. (c) The dynamic phase diagram for domain dynamics at finite $T$. The (red) line $E_1$ and (blue) line $E_2$ separate no nuclei, creep, and flow regimes.

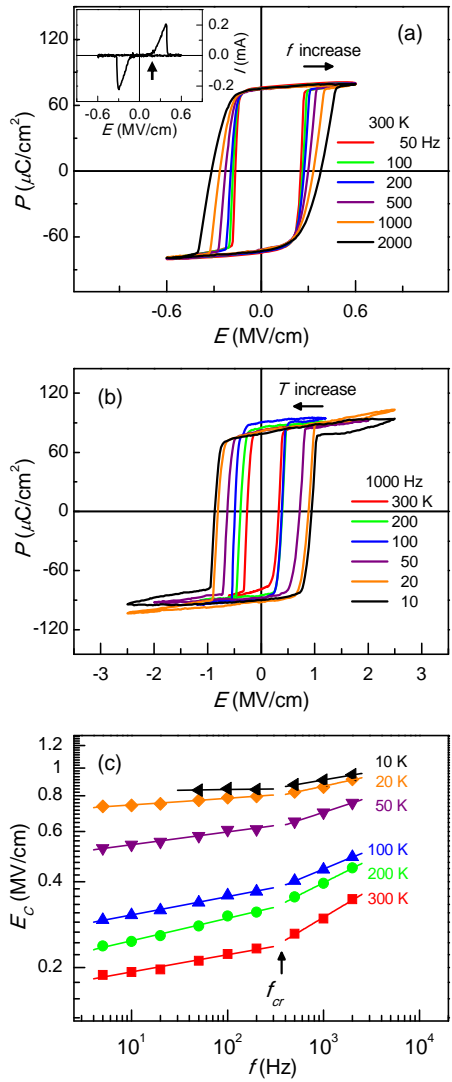

Fig. 1. S. M. Yang *et al*.

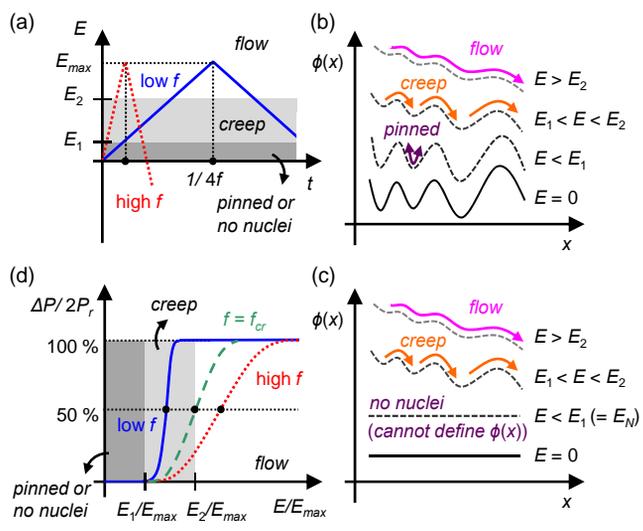

Fig. 2. S. M. Yang *et al*.

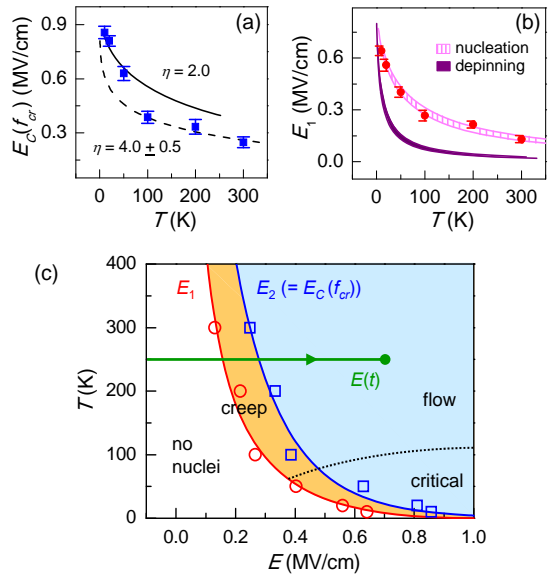

Fig. 3. S. M. Yang *et al.*